\documentclass[a4paper,twoside]{article}

\usepackage[latin1]{inputenc}
\usepackage{hyperref}

\usepackage[ngerman]{babel}
\usepackage{bibgerm}
\usepackage{a4wide}

\usepackage{fancyhdr}
\usepackage{graphicx}
\usepackage{mathpaper}
\usepackage{url}
\usepackage{color}

\renewcommand \secref[1]{Abschnitt~\ref{sec:#1}}     
\renewcommand \figref[1]{Abbildung~\ref{fig:#1}}     

\newcommand\figcaption[2]{\caption{\textit{#2}}\figlabel{#1}}

\newcommand\todo[1]{}

\author{
	Henning Thielemann \\
	\\
	Institut für Informatik \\
	Martin-Luther-Universität Halle-Wittenberg \\
	Von-Seckendorff-Platz 1 \\
	06122 Halle \\
	henning.thielemann@informatik.uni-halle.de
}
\title{Live-Musikprogrammierung in Haskell}
\begin{document}
\maketitle

\begin{abstract}
Ziel unserer Arbeit ist es,
algorithmische Musik interaktiv und mit mehreren Teilnehmern zu komponieren.
Dazu entwickeln wir einen Interpreter
für eine Teilsprache der nicht-strikten funktionalen Programmiersprache Haskell~98,
der es erlaubt, das Programm noch während seiner Ausführung zu ändern.
Unser System eignet sich sowohl für den Live-Einsatz zur Musikprogrammierung
als auch als Demonstrations- und Lernumgebung für funktionale Programmierung.
\end{abstract}

\section{Einführung}

Unser Ziel ist es, Musik durch Algorithmen zu beschreiben.
Wir wollen Musik nicht wie auf dem Notenblatt
als mehr oder weniger zusammenhanglose Folge von Noten darstellen,
sondern wir wollen Strukturen ausdrücken.
Beispielsweise wollen wir nicht die einzelnen Noten einer Begleitung aufschreiben,
sondern die Begleitung durch ein allgemeines Muster
und die Folge von Harmonien ausdrücken.
Als weiteres Beispiel mag ein Komponist dienen,
der eine Folge von zufälligen Noten verwenden möchte.
Er möchte die Noten aber nicht einzeln aufschreiben,
sondern die Idee ausdrücken, dass es eine zufällige Folge von Noten ist.
Dem Interpreten wäre es damit freigestellt,
eine andere aber ebenso zufällige Folge von Noten zu spielen.

Der Programmierer soll den Grad der Strukturierung frei wählen können.
Beispielsweise soll es möglich sein,
"`von Hand"' eine Melodie zu komponieren,
diese mit einem Tonmuster zu begleiten,
für das lediglich eine Folge von Harmonien vorgegeben wird,
und das ganze mit einem
vollständig automatisch berechneten Rhythmus zu unterlegen.

Mit dem bewussten Abstrahieren von der tatsächlichen Musik
wird es aber schwierig,
beim Programmieren das Ergebnis der Komposition abzuschätzen.
Bei Musik, die nicht streng nach Takten und Stimmen organisiert ist,
fällt es schwerer, einen bestimmten Zeitabschnitt
oder eine bestimmte Auswahl an Stimmen zur Probe anzuhören.
Auch der klassische Zyklus von
"`Programm editieren, Programm prüfen und übersetzen, Programm neu starten"'
steht dem kreativen Ausprobieren entgegen.
Selbst wenn Übersetzung und Neustart sehr schnell abgeschlossen sind,
so muss doch das Musik erzeugende Programm und damit die Musik abgebrochen
und neu begonnen werden.
Insbesondere beim gemeinsamen Spiel mit anderen Musikern
ist das nicht akzeptabel.

In unserem Ansatz verwenden wir zur Musikprogrammierung
eine rein funktionale Programmiersprache mit Bedarfsauswertung
\cite{hughes1984fpmatter},
die nahezu eine Teilsprache von Haskell~98 \cite{peyton-jones1998haskell} ist.
Unsere Beiträge zur interaktiven Musikprogrammierung sind
Konzepte und ein lauffähiges System,
welches folgendes bieten:
\begin{itemize}
\item Algorithmische Musikkomposition,
bei der das Programm verändert werden kann, während die Musik läuft
(\secref{live-coding}),
\item gleichzeitige Arbeit mehrerer Programmierer an einem Musikstück
unter Anleitung eines "`Dirigenten"'
(\secref{multiuser}).
\end{itemize}



\section{Funktionale Live-Programmierung}


\subsection{Live-Coding}
\seclabel{live-coding}

Wir wollen Musik ausgeben als eine Liste von MIDI-Ereignissen
\cite{mma1996midi},
also Ereignissen der Art
"`Klaviaturtaste gedrückt"', "`Taste losgelassen"',
"`Instrument gewechselt"', "`Klangregler verändert"'
und Warte-Anweisungen.
Ein Ton mit Tonhöhe C-5, einer Dauer von 100 Millisekunden
und einer normalen Intensität
soll geschrieben werden als:
\begin{verbatim}
main =
  [ Event (On c5 normalVelocity)
  , Wait 100
  , Event (Off c5 normalVelocity)
  ] ;

c5 = 60 ;
normalVelocity = 64 ;
.
\end{verbatim}
Mit der Listenverkettung "`\verb|++|"' lässt sich damit bereits
eine einfache Melodie beschreiben.
\begin{verbatim}
main =
   note qn c ++ note qn d ++ note qn e ++ note qn f ++
   note hn g ++ note hn g ;

note duration pitch =
  [ Event (On pitch normalVelocity)
  , Wait duration
  , Event (Off pitch normalVelocity)
  ] ;

qn = 200 ;  -- quarter note - Viertelnote
hn = 2*qn ; -- half note    - halbe Note

c = 60 ;
d = 62 ;
e = 64 ;
f = 65 ;
g = 67 ;
normalVelocity = 64 ;
\end{verbatim}
Diese Melodie lässt sich endlos wiederholen,
indem wir am Ende der Melodie wieder mit dem Anfang fortsetzen.
\begin{verbatim}
main =
   note qn c ++ note qn d ++ note qn e ++ note qn f ++
   note hn g ++ note hn g ++ main ;
\end{verbatim}
Die so definierte Liste \verb|main| ist unendlich lang,
lässt sich aber mit der Bedarfsauswertung schrittweise berechnen
und an einen MIDI-Synthesizer senden.
Dank der Bedarfsauswertung kann man die Musik
als reine Liste von Ereignissen beschreiben.
Das Programm muss und kann selbst keine Ausgabebefehle ausführen.
Der Versand der MIDI-Kommandos wird vom Interpreter übernommen.

In einem herkömmlichen interaktiven Interpreter%
\footnote{Der Interpreter wäre hier im wahrsten Sinne des Wortes
 der musikalische Interpret}
wie dem \verb|GHCi|
würde man die Musik etwa so wiedergeben:
\begin{verbatim}
Prelude> playMidi main       .
\end{verbatim}
Will man die Melodie ändern, müsste man die Musik beenden
und die neue Melodie von vorne beginnen.
Wir wollen aber die Melodie ändern,
während die alte Melodie weiterläuft,
und dann die alte Melodie nahtlos in die neue übergehen lassen.
Mit anderen Worten:
Der aktuelle Zustand des Interpreters setzt sich zusammen
aus dem Programm und dem Zustand der Ausführung.
Wir wollen das Programm austauschen,
aber den Zustand der Ausführung beibehalten.
Das bedeutet, dass der Zustand in einer Form gespeichert sein muss,
der auch nach Austausch des Programms einen Sinn ergibt.

Wir lösen dieses Problem wie folgt:
Der Interpreter betrachtet das Programm als Menge von Termersetzungsregeln
und die Ausführung des Programms besteht darin,
die Ersetzungsregeln wiederholt anzuwenden,
solange bis der Startterm \verb|main| so weit reduziert ist,
dass die Wurzel des Operatorbaums ein Terminalsymbol
(hier: ein Datenkonstruktor) ist.
\todo{Konstruktor erklären?}
Für die musikalische Verarbeitung testet der Interpreter weiterhin,
ob die Wurzel ein Listenkonstruktor ist
und falls es eine nichtleere Liste ist,
reduziert er das führende Listenelement vollständig
und prüft, ob es ein MIDI-Ereignis darstellt.
Ausführungszustand des Interpreters ist der reduzierte Ausdruck.
Während der Interpreter die vorletzte Note
in der Schleife des obigen Programms wiedergibt,
wäre dies beispielsweise:
\begin{verbatim}
Wait 200 :
   (Event (Off g normalVelocity) : (note hn g ++ main))
.
\end{verbatim}

Der Ausdruck wird immer so wenig wie möglich reduziert,
gerade so weit, dass das nächste MIDI-Ereignis bestimmt werden kann.
Das erlaubt es zum einen, eine unendliche Liste wie \verb|main| zu verarbeiten
und zum anderen führt es dazu,
dass in dem aktuellen Term wie oben angegeben,
noch die Struktur des restlichen Musikstücks zu erkennen ist.
Der abschließende Aufruf von \verb|main| ist beispielsweise noch vorhanden.
Wenn wir jetzt die Definition von \verb|main| ändern,
wird diese veränderte Definition verwendet,
sobald \verb|main| reduziert wird.
Wir können auf diese Weise die Melodie innerhalb der Wiederholung ändern,
beispielsweise so:
\begin{verbatim}
main =
   note qn c ++ note qn d ++ note qn e ++ note qn f ++
   note qn g ++ note qn e ++ note hn g ++ main ;
.
\end{verbatim}
Wir können aber auch folgende Änderung vornehmen
\begin{verbatim}
main =
   note qn c ++ note qn d ++ note qn e ++ note qn f ++
   note hn g ++ note hn g ++ loopA ;
\end{verbatim}
und damit erreichen,
dass nach einer weiteren Wiederholung der Melodie
die Musik mit einem Abschnitt namens \verb|loopA| fortgesetzt wird.

Wir halten an dieser Stelle fest,
dass sich die Bedeutung eines Ausdrucks während des Programmablaufs ändern kann.
Damit geben wir eine wichtige Eigenschaft der rein funktionalen Programmierung auf.
Wenn wir von Live-Änderungen Gebrauch machen,
ist unser System also nicht mehr "`referential transparent"'.
Beispielsweise hätten wir die ursprüngliche Schleife
auch mit der Funktion \verb|cycle| implementieren können
\begin{verbatim}
main =
  cycle
    ( note qn c ++ note qn d ++ note qn e ++ note qn f ++
      note hn g ++ note hn g ) ;
\end{verbatim}
und wenn \verb|cycle| definiert ist als
\begin{verbatim}
cycle xs = xs ++ cycle xs ;
\end{verbatim}
dann würde dies reduziert werden zu
\begin{verbatim}
( note qn c ++ note qn d ++ note qn e ++ note qn f ++
  note hn g ++ note hn g )
++
cycle
   ( note qn c ++ note qn d ++ note qn e ++ note qn f ++
     note hn g ++ note hn g ) ;
.
\end{verbatim}
Die Schleife könnte dann nur noch verlassen werden,
wenn man die Definition von \verb|cycle| ändert.
Diese Änderung würde aber alle Aufrufe von \verb|cycle|
im aktuellen Term gleichermaßen betreffen.
Zudem wäre es bei einem strengen Modulsystem ohne Importzyklen unmöglich,
im Basis-Modul \verb|List|, in dem \verb|cycle| definiert ist,
auf Funktionen im Hauptprogramm zuzugreifen.
Dies wäre aber nötig, um die \verb|cycle|-Schleife nicht nur verlassen,
sondern auch im Hauptprogramm fortsetzen zu können.

Wir erkennen an diesem Beispiel,
dass es vorausschauend besser sein kann,
mit einer "`von Hand"' programmierten Schleife
der Form \verb|main = ... ++ main|
eine Sollbruchstelle zu schaffen,
an der man später neuen Code einfügen kann.

Neben der seriellen Verkettung von musikalischen Ereignissen
benötigen wir noch die parallele Komposition,
also die simultane Wiedergabe von Melodien, Rhythmen usw.
Auf der Ebene der MIDI-Kommandos bedeutet dies,
dass die Kommandos zweier Listen geeignet miteinander verzahnt werden müssen.
Wir wollen die Definition der entsprechenden Funktion "`\verb|=:=|"'
der Vollständigkeit halber hier wiedergeben.
\begin{verbatim}
(Wait a : xs) =:= (Wait b : ys) =
  mergeWait (a<b) (a-b) a xs b ys ;
(Wait a : xs) =:= (y : ys) =
  y : ((Wait a : xs) =:= ys) ;
(x : xs) =:= ys  =  x : (xs =:= ys) ;
[] =:= ys  =  ys ;

mergeWait _eq 0 a xs _b ys =
  Wait a : (xs =:= ys) ;
mergeWait True d a xs _b ys =
  Wait a : (xs =:= (Wait (negate d) : ys)) ;
mergeWait False d _a xs b ys =
  Wait b : ((Wait d : xs) =:= ys) ;
\end{verbatim}

Die grafische Bedienschnittstelle unseres Systems
ist in \figref{bildschirm} zu sehen.
Im linken oberen Teil kann der Benutzer den Programmtext eingeben.
Mit einer bestimmten Tastenkombination
kann er den Programmtext auf Syntaxfehler prüfen
und in den Programmspeicher des Interpreters übernehmen.
Das vom Interpreter ausgeführte Programm ist im rechten oberen Teil zu sehen.
In diesem Teil hebt der Interpreter außerdem hervor,
welche Teilausdrücke reduziert werden mussten,
um den vorhergehenden in den aktuellen Ausdruck zu überführen.
Auf diese Weise kann man den Verlauf der Melodie visuell verfolgen.
Der aktuelle Term des Interpreters ist im unteren Teil
der Bedienoberfläche dargestellt.
Die in der Abbildung wiedergegebenen Texte entsprechen
im Wesentlichen unserem einführenden Beispiel,
weisen aber unsere Melodie zusätzlich noch einem MIDI-Kanal zu
und verwenden Definitionen von \verb|++| und \verb|map|,
welche auf \verb|foldr| aufbauen.
\begin{figure}[htb]
  \begin{center}
    \includegraphics[width=\hsize]{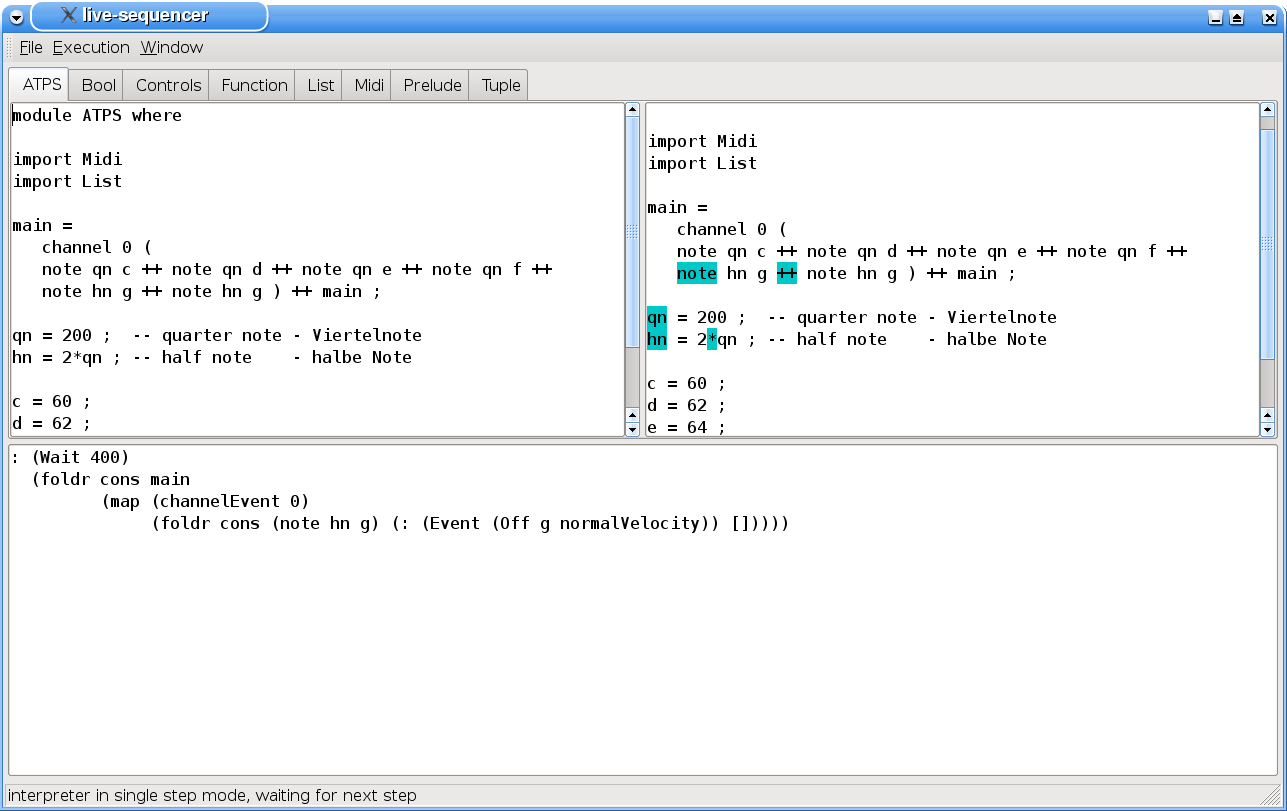}
    \figcaption{bildschirm}{Der Interpreter im Betrieb}
  \end{center}
\end{figure}

Das System verfügt über die Ausführungsmodi
"`Echtzeit"', "`Zeitlupe"' und "`Einzelschritt"'.
Der Echtzeit-Modus gibt die Musik wieder,
so wie es die Notenlängen erfordern,
während die anderen beiden Modi
die Warte-Ereignisse ignorieren
und stattdessen nach jedem Element der MIDI-Kommandoliste
eine Pause machen.
Die beiden letzten Modi sind
zur Beobachtung der Ausführung und zur Fehlersuche gedacht.
Sie eignen sich auch für den Einsatz in der Lehre,
zur Erläuterung, wie ein Interpreter einer funktionalen Sprache
mit Bedarfsauswertung im Prinzip funktioniert.

Der Interpreter ist in Haskell
mit dem Glasgow Haskell Compiler GHC \cite{ghc2012} implementiert
und verwendet WxWidgets \cite{smart2011wxwidgets}
für die grafische Bedienoberfläche.
Die verarbeitete Sprache unterstützt
"`Pattern matching"', vordefinierte Infix-Operatoren,
Funktionen höherer Ordnung, unvollständige Funktionsanwendung.
Aus Gründen der einfachen Implementierung
gibt es bislang folgende Einschränkungen:
Die interpretierte Sprache ist dynamisch typisiert,
und kennt als Objekte ganze Zahlen, Texte und Konstruktoren.
Sie ist formatfrei,
weswegen nach Deklarationen stets Semikola gesetzt werden müssen.
Viele syntaktische Besonderheiten werden nicht unterstützt,
beispielsweise "`List Comprehension"', "`Operator Section"', Do-Notation,
Let- und Case-Notation, frei definierbare Infix-Operatoren.
Ein- und Ausgabeoperationen sind ebenfalls nicht verfügbar.

\subsection{Verteilte Programmierung}
\seclabel{multiuser}

Unser System soll es auch erlauben,
das Publikum in eine Aufführung
oder Studenten in die Vorlesung
durch Programmieren einzubeziehen.
Die typische Situation dafür ist,
dass der Vortragende die Bedienoberfläche des Programms
an die Wand projiziert,
die Zuhörer die erzeugte Musik über eine Musikanlage hören
und dass die Zuhörer über ein Funknetz und einen Browser
Kontakt mit dem Vortragsrechner aufnehmen können.

Die implementierte funktionale Sprache
verfügt über ein einfaches Modulsystem.
Der Vortragende kann auf diese Weise
ein Musikstück in mehrere Abschnitte oder Tonspuren zerlegen,
und jeden dieser Teile in einem Modul ablegen.
Die Module wiederum kann er Zuhörern zuweisen.
Außerdem kann der Vortragende durch Einfügen eines bestimmten Kommentars festlegen,
ab welcher Zeile der Zuhörer den Modulinhalt verändern darf.
In den Zeilen davor stehen üblicherweise der Modulname,
die Liste der exportierten Bezeichner,
die Liste der importierten Module
und grundlegende Definitionen.
Auf diese Weise kann der Vortragende eine Schnittstelle
für jedes Modul vorgeben.

Ein Zuhörer kann nun über einen WWW-Browser
ein Modul abrufen und den ver\-än\-der\-ba\-ren Teil editieren.
(Siehe \figref{browser})
Nach dem Editieren kann er den veränderten Inhalt an den Server schicken.
Dieser ersetzt im Editor den Modultext unterhalb des Markierungskommentars
mit dem neuen Inhalt.
Dann wird der Text syntaktisch geprüft und
im Erfolgsfall an den Interpreter weitergeleitet.
Ist der Text syntaktisch nicht korrekt, so bleibt er im Editor,
damit er notfalls vom Vortragenden überprüft und korrigiert werden kann.
\begin{figure}[htb]
  \begin{center}
    \includegraphics[width=\hsize]{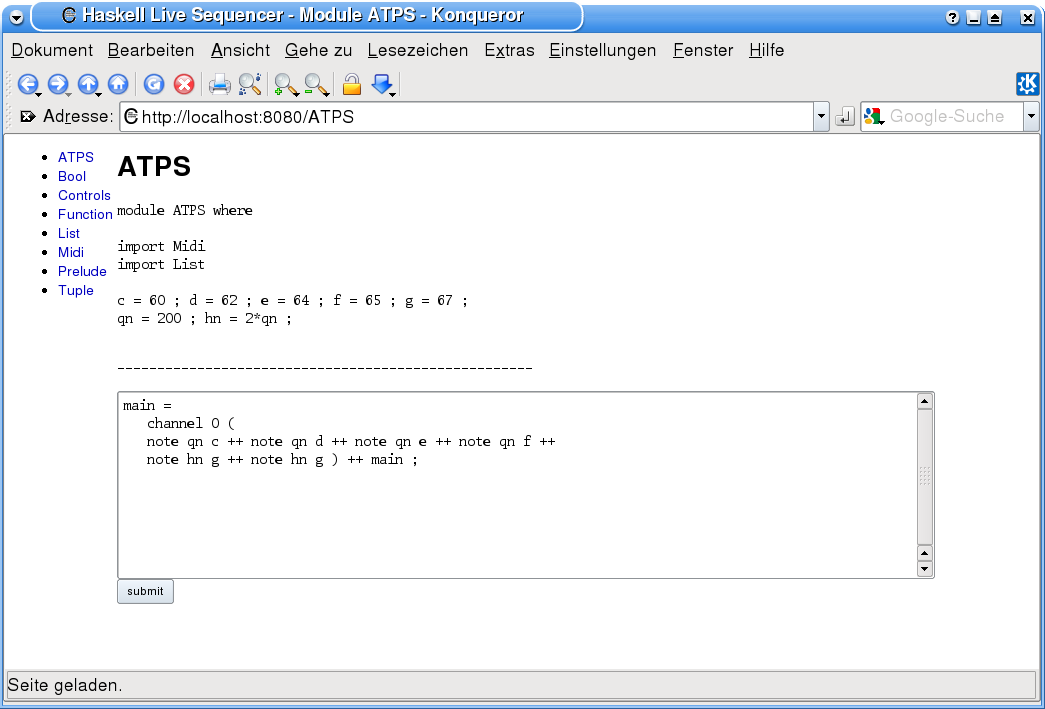}
    \figcaption{browser}{Zugriff auf ein Modul über das Netz}
  \end{center}
\end{figure}

Im Allgemeinen wird es nicht gelingen,
ohne Vorbereitung auf diese Weise ein völlig neues Musikstück zu erschaffen.
Der Vortragende kann aber seine Aufführung vorbereiten,
indem er sich eine Aufteilung in Module überlegt
und diese mit grundlegenden Definitionen füllt.
Dies können Definitionen von Funktionen sein,
die eine Liste von Nullen und Einsen in einen Rhythmus verwandeln,
oder eine Liste von Zahlen in ein Akkordmuster oder eine Bassbegleitung.
Der Vortragende kann dann durch Vorgabe von Takt und von Harmonien sicher stellen,
dass die einzelnen Stücke zusammenpassen.
Der Vortragende übernimmt in diesem Szenario nicht mehr die Rolle des Komponisten,
sondern eher die Rolle des Dirigenten.


\section{Verwandte Arbeiten}

Algorithmische Komposition hat inzwischen eine lange Tradition.
Als Beispiele seien hier nur Mozarts musikalische Würfelspiele
und die Illiac-Suite \cite{hiller1959illiacsuite} genannt.
Auch gibt es mit Haskore \cite{hudak1996haskore}
seit einiger Zeit die Mög\-lich\-keit,
Musik in Haskell zu programmieren
und damit verschiedene Klangerzeuger über MIDI zu steuern
oder mit CSound, SuperCollider
oder mit in Haskell geschriebenen Audiosynthesefunktionen
Audiodateien zu erzeugen.
Haskore baut ebenfalls auf der Bedarfsauswertung auf
und erlaubt die elegante Definition
von formal großen oder unendlichen Musikstücken
bei geringem tatsächlichem Speicherverbrauch bei der Interpretation.
Das kreative Komponieren wird allerdings dadurch erschwert,
dass man Änderungen erst nach Abbruch und Neustart des Programms hören kann.

Ein sehr populärer Ansatz zur Programmierung
von Animationen, Robotersteuerungen, grafischen Bedienschnittstellen
und Audiosignalverarbeitung in Haskell
ist die funktionale reaktive Programmierung
\cite{elliott1997fran}.
Wie für unsere Musikstücke wird auch hier der zeitliche Verlauf
einer Animation, einer graphischen Bedienoberfläche oder ähnlichem
durch eine Art unendliche Liste beschrieben.
Darüber hinaus soll ein FRP-Programm auch auf äußere Ereignisse,
wie zum Beispiel Bewegungen der Computermaus reagieren können.
Im Gegensatz zu unserer Arbeit ist bisher allerdings nicht möglich,
ein FRP-Programm während seiner Ausführung zu ändern.

Eine funktionale, aber nicht rein funktionale, Programmiersprache,
die Än\-de\-run\-gen des Programmtextes während der Programmausführung erlaubt,
ist Erlang
\cite{armstrong1997erlang}.
Erlang folgt der strengen Semantik.
Man könnte in Erlang eine Folge von MIDI-Kommandos
nicht durch eine (lazy) Liste von Konstruktoren beschreiben,
sondern bräuchte Iteratoren oder ähnliches.
In ein laufendes Erlang-Programm kann neuer Programmcode
auf zwei verschiedene Weisen eingebracht werden,
entweder in dem das laufende Programm Funktionen
(zum Beispiel Lambda-Ausdrücke) aufruft,
die ihm als Nachrichten zugesandt werden
oder indem ein Erlang-Modul durch ein neues Modul ersetzt wird.
Wird ein Erlang-Modul nachgeladen,
so behält das Laufzeitsystem die alte Version des Moduls im Speicher,
um laufende Programmteile ausführen zu können.
Lediglich modul-externe Aufrufe
springen in das neue Modul,
wobei man einen externen Aufruf auch
innerhalb desselben Moduls starten kann.
Auch in Erlang müssen also Sollbruchstellen
(externe Aufrufe oder Funktionsübernahme aus Nachrichten)
zum späteren Einfügen von neuem Code geschaffen werden.

Unser Ansatz zur Änderung von Programmtext während der Programmausführung
ist damit sehr ähnlich zum "`Hot Code loading"' in Erlang.
Die Bedarfsauswertung in unserem Interpreter führt jedoch dazu,
dass beträchtliche Teile des Programmtextes im aktuellen Term enthalten sind.
Diese werden von Änderungen an einem Modul nicht unmittelbar betroffen.
Auf diese Weise ist es in unserem Ansatz nicht unbedingt nötig,
zwei Versionen eines Moduls im Speicher zu halten,
um einen glatten Übergang von altem zu neuem Programmtext zu erreichen.

\todo{Java-Code-Änderung in Eclipse-Debugger?}

Sogenanntes Musik-Live-Coding,
also das Programmieren eines musikerzeugenden Programms,
während die Musik läuft,
war bislang Spezialsprachen
wie ChucK \cite{wang2004chuck} und
SuperCollider/SCLang \cite{mccartney1996supercollider}
und ihren Implementierungen vorbehalten.
Diese Sprachen sind beim Kontrollfluss
an das imperative Programmierparadigma angelehnt
und beim Typsystem an die objektorientierte Programmierung.
Im wesentlichen funktionieren beide wie eine Client-Server-Lösung,
wobei der Server die Klänge erzeugt
und parallel zu einem Kommandozeileninterpreter läuft,
von dem aus man Befehle an den Server schicken kann.

Auch in unserer Architektur läuft die Klangerzeugung
parallel zur eigentlichen Programmierung
und wird mit (MIDI-)Kommandos gesteuert.
Jedoch wird in unserem Ansatz nicht programmiert,
wie sich die Klangerzeugung ändern soll,
sondern das erzeugende Programm wird direkt geändert.



\section{Folgerungen und zukünftige Arbeiten}

Unsere vorgestellte Technik zeigt einen neuen Weg
der Live-Programmierung von Musik auf,
der sich möglicherweise auch auf die Wartung
anderer lange laufender funktionaler Programme übertragen lässt.
Dennoch zeigt sich, dass man schon beim Programmieren
einer ersten Version gewisse Sollbruchstellen vorsehen muss,
an denen man später im laufenden Programm Änderungen einfügen kann.
Auch mit automatischen Optimierungen des Programms
müssen wir jetzt vorsichtig sein,
denn eine Optimierung könnte eine solche Sollbruchstelle entfernen.
Wenn ein Programm zur Laufzeit geändert wird,
so sind Funktionen eben nicht mehr "`referential transparent"',
womit wir eine wichtige Eigenschaft der funktionalen Programmierung aufgeben.

\paragraph{Typsystem}
Um die Gefahr zu verringern, dass ein Musikprogramm
nach einer Änderung wegen eines Programmfehlers abbricht,
ist der nahe liegende nächste Schritt
der Einsatz eines statischen Typprüfers.
Dieser müsste nicht nur testen,
ob das vollständige Programm nach Austausch eines Moduls noch typkorrekt ist,
sondern er müsste zudem testen,
dass der aktuelle Term im Interpreter
bezüglich des neuen Programms typkorrekt ist.

Noch wichtiger wird ein Typprüfer im Mehrbenutzerbetrieb.
Der Leiter einer Programmierveranstaltung mit mehreren Programmierern
könnte jedem Teilnehmer Typsignaturen
im nicht editierbaren Bereich seines Moduls vorgeben,
die der Teilnehmer implementieren muss.
Der Typprüfer würde dafür sorgen,
dass Teilnehmer nur Änderungen einschicken können,
die zum Rest des Musikstücks passen.

\paragraph{Auswertungsstrategie}
Derzeit ist unser Interpreter sehr einfach gehalten.
Der aktuelle Term ist ein reiner Baum.
In dieser Darstellung können wir nicht ausdrücken,
dass der Wert eines Terms mehrmals verwendet wird.
(Es gibt also kein "`sharing"'.)
Wenn beispielsweise \verb|f| definiert ist als \verb|f x = x:x:[]|,
dann wird der Aufruf \verb|f (2+3)| reduziert zu \verb|(2+3) : (2+3) : []|.
Wenn weiterhin das erste Listenelement zu \verb|5| reduziert wird,
wird das zweite Listenelement nicht reduziert.
Wir erhalten also \verb|5 : (2+3) : []| und nicht \verb|5 : 5 : []|.
Da der Term nur ein Baum ist und kein Graph,
brauchen wir keine eigene Speicherverwaltung,
sondern können uns auf die automatische Speicherverwaltung
des GHC-Laufzeitsystems verlassen,
in dem der Interpreter läuft.
Wenn ein Teilterm nicht mehr benötigt wird,
so wird er aus dem Operatorbaum entfernt
und früher oder später vom Laufzeitsystem des GHC freigegeben.

Selbst einfache korekursive Definitionen
wie die der Fibonacci-Zahlen durch
\begin{verbatim}
fix (\fibs -> 0 : 1 : zipWith (+) fibs (tail fibs))
\end{verbatim}
führen bei diesem Auswertungsverfahren zu einem unbegrenzten Anstieg der Termgröße.
In Zukunft sollen daher weitere Auswertungsstrategien
wie zum Beispiel die Graphreduktion mit der STG-Maschine
\cite{peyton-jones1992stg}
hinzukommen,
die dieses und weitere Probleme lösen.
Anstelle eines Operatorbaums würde der aktuelle Term
dann aus einem Operatorgraph bestehen,
die Anwendung der Funktionsdefinitionen
und damit die Möglichkeit der Live-Änderung einer Definition
bliebe prinzipiell erhalten.
Die Gefahr bei Live-Mu\-sik\-pro\-gram\-mie\-rung liegt natürlich darin,
dass Programmänderungen abhängig von der Auswertungsstrategie
verschiedene Auswirkungen auf den Programmablauf haben können.
Das Verwenden des gleichen Objektes im Speicher
an verschiedenen Stellen im aktuellen Term ("`sharing"')
würde zwar im obigen Beispiel der Fibonacci-Zahlen
den Speicherverbrauch begrenzen,
könnte aber auch verhindern,
dass eine geänderte Definition der aufgerufenen Funktionen
noch berücksichtigt wird.
Der Einzelschrittmodus würde es ermöglichen,
in der Lehre verschiedene Auswertungsverfahren zu demonstrieren
und Vor- und Nachteile miteinander zu vergleichen.

Offen ist, ob und wie wir unser System,
das Änderungen des Programms während des Programmablaufs zulässt,
direkt in eine existierende Sprache wie Haskell einbetten können.
Dies würde es uns vereinfachen,
die Wechselwirkung zwischen Programmänderungen,
Optimierungen und Auswertungsstrategien zu untersuchen.


\paragraph{Hervorhebungen}
Es gibt noch ein weiteres interessantes offenes Problem:
Wie kann man Textstellen im Programm passend zur erzeugten Musik hervorheben?
Es liegt nahe, die jeweils gespielten Noten hervorzuheben.
Dies wird zur Zeit dadurch erreicht,
dass in einer Wartephase alle die Symbole hervorgehoben werden,
welche seit der letzten Wartephase durch den Interpreter reduziert wurden.
Wenn aber eine langsame Melodie
parallel zu einer schnellen Folge von Regleränderungen abgespielt wird,
so führt das dazu, dass die Noten der Melodie nur kurz hervorgehoben werden,
nämlich immer nur für die kurze Zeit, in der der Reglerwert konstant bleibt.
Wir würden aber erwarten,
dass die Hervorhebung eines Musikteils
nicht von parallel laufenden Teilen beeinflusst wird.
Formal könnten wir es so ausdrücken:
Gegeben seien die serielle Komposition \verb|++| und
die parallele Komposition \verb|=:=|,
die sowohl für Terme als auch für Hervorhebungen definiert sein sollen.
Gegeben sei weiterhin die Abbildung \verb|highlight|,
welche einen Term seiner Visualisierung zuordnet.
Dann soll für zwei beliebige Musikobjekte \verb|a| und \verb|b| gelten:
\begin{verbatim}
highlight (a ++  b) = highlight a ++  highlight b
highlight (a =:= b) = highlight a =:= highlight b
\end{verbatim}
Wenn man alle Symbole hervorhebt,
die mittelbar an der Erzeugung
eines \verb|NoteOn|- oder \verb|NoteOff|-MIDI-Kommandos beteiligt waren,
dann erhält man eine Funktion \texttt{high\-light} mit diesen Eigenschaften.
Allerdings führt sie dazu, dass die Notenaufrufe kumulativ hervorgehoben werden.
In
\begin{verbatim}
note qn c ++ note qn d ++ note qn e ++ note qn f
\end{verbatim}
werden bei Wiedergabe von \verb|note qn e|
auch \verb|note qn c| und \verb|note qn d| hervorgehoben,
denn diese erzeugen Listen und dass diese Listen endlich sind,
ist ein Grund dafür, dass aktuell \verb|note qn e| wiedergegeben werden kann.
Die Aufrufe \verb|note qn c| und \verb|note qn d|
sind also notwendigerweise daran beteiligt,
dass \verb|note qn e| reduziert werden kann.


\paragraph{Zeitsteuerung}
Eine weitere Schwierigkeit besteht
im zeitlich präzisen Versand der MIDI-Kom\-man\-dos.
Bislang wartet der Interpreter bis zu dem Zeitpunkt,
an dem eine MIDI-Nachricht verschickt werden soll,
und beginnt dann erst mit der Berechnung
des entsprechenden Listenelements.
Wir vertrauen also darauf,
dass die Berechnung schnell genug beendet wird
und sich der Versand nicht allzu stark verzögert.
Bei komplizierteren Berechnungen trifft diese Annahme natürlich nicht zu.
Eine höhere Präzision könnten wir erreichen,
indem wir die MIDI-Nachrichten mit einem Zeitstempel versehen
und einige Zeit im Voraus verschicken.
Das wirft neue Probleme der Synchronisation von Musik
und grafischer Darstellung des Interpreterzustandes auf
und es würde auch heißen,
dass die Musik erst verzögert angehalten werden kann
und man nur verzögert in einen anderen Ausführungsmodus wechseln kann.


\section{Danksagungen}

Dieses Projekt basiert auf einer Idee von Johannes Waldmann,
die ich gemeinsam mit ihm entwickelt und in einen Prototypen umgesetzt habe.
Ich möchte mich bei ihm, Heinrich Apfelmus und den anonymen Gutachtern
für das gründliche Lesen und
für zahlreiche Verbesserungsvorschläge zu diesem Artikel bedanken.

Informationen über dieses Projekt
zu Programmentwicklung, Demonstrationsvideos und Artikeln
können Sie unter
\begin{quote}
\url{http://www.haskell.org/haskellwiki/Live-Sequencer}
\end{quote}
abrufen.

\bibliography{thielemann,audio,haskell,literature}

\end{document}